# Possible Nonplanar Structure of Phagraphene

# and Its Thermal Stability

A.I. Podlivaev, L.A. Openov

National Research Nuclear University "MEPhI", Kashirskoe sh. 31, Moscow 115409,

Russian Federation

E-mail addresses: LAOpenov@mephi.ru, laopenov@gmail.com

## ABSTRACT

It is shown that phagraphene, a recently predicted planar allotrope of graphene with Dirac fermions, is unstable or, at least, almost unstable with respect to transverse atomic displacements. This result is obtained by numerical calculations in the framework of both the tight-binding model and the density functional theory. A nonplanar atomic configuration of phagraphene has a wavy shape and is almost degenerate in energy with the planar configuration. The main types of possible structural defects in phagraphene are determined. The temperature dependence of characteristic times of their formation is found.

The discovery of graphene [1], a hexagonal monolayer of carbon atoms with unique electronic [2] and mechanical [3] characteristics, has stimulated experimental and theoretical studies aimed at a search for novel quasi-two-dimensional carbon-based materials [4–10]. Graphene allotropes where the dispersion curves for electrons and holes have a conic shape and touch each other are of a special interest [9, 11–13]. Indeed, in such a case, the charge carriers are Dirac fermions with a very high mobility. The hexagonal symmetry is not a mandatory requirement for the existence of Dirac cones [12, 13]; therefore, the number of such allotropes turns out to be quite large. However, the structure of most of them (at least of all graphynes [11–13]) include carbon chains of different lengths; i.e., the electron orbitals for an appreciable part of atoms are sp hybridized. This results in the weakening of interatomic bonds (in comparison to those between $sp^2$ hybridized atoms in graphene) and in the corresponding lowering of stability (an increase in energy).

Numerical calculations performed in [14] in the framework of both the tight-binding model and the density functional theory (DFT) predicted the existence of a new planar allotropic form of graphene with Dirac cones, whose stability is only slightly below that of graphene. The authors of [14] call it *phagraphene*. In addition to six-atomic graphene rings formed by C–C bonds, phagraphene includes five- and seven-atomic rings (see Fig. 1). All of them are packed in such a way that each atom has three nearest neighbors, as occurs in graphene. Therefore, the electron orbitals are $sp^2$ hybridized. The binding energy $E_b = -9.03$ eV/atom exceeds that of graphene, $E_b = -9.23$ eV/atom, only by $\Delta E_b = 0.20$ eV/atom, whereas $\Delta E_b = 0.72$ eV/atom and more for other allotropes with the Dirac cones (here, we should mention that the DFT significantly overestimates the absolute values of $E_b$ in comparison with the experimental data but allows at the same time determining $\Delta E_b$ with a high accuracy)

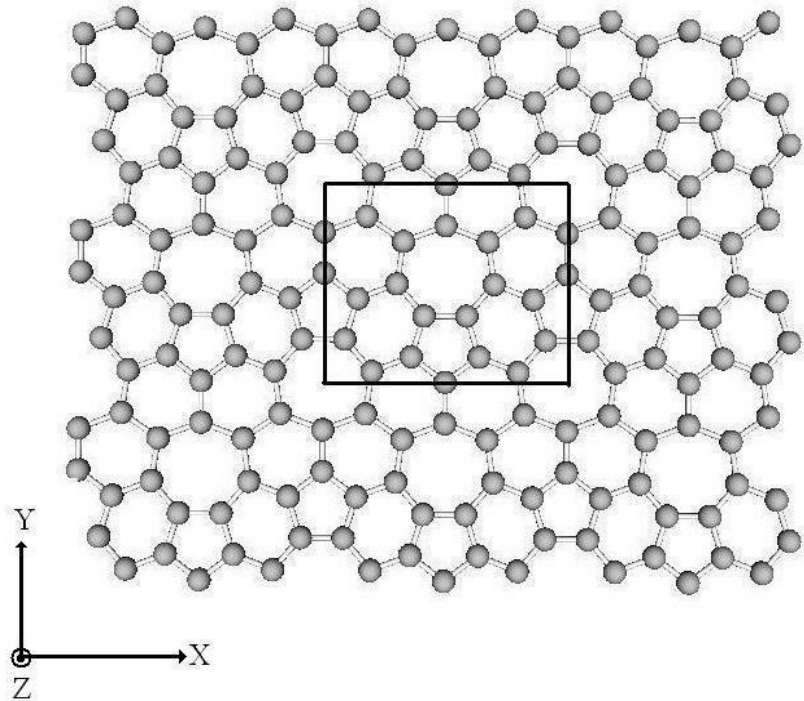

Fig. 1. Planar 3×3 supercell of phagraphene. The 20-atomic unit cell is shown in the box.

While calculating the phonon spectrum of phagraphene, the authors of [14] did not find there any imaginary frequencies. Therefore, they concluded that phagraphene is dynamically stable; i.e., its atomic configuration corresponds to the local minimum of the potential energy surface (PES). This result seems to be rather doubtful, to say the least. The fact is that phagraphene, being a quasi-two-dimensional object, exists in three-dimensional space. The previous analysis of the Stone–Wales defects [15] in graphene demonstrated that the neighborhood of five-, six-, and seven-atomic rings leads to the instability of the strictly planar structure of this defect with respect to the transverse atomic displacements [16]. This means that the planar configuration corresponds to the saddle point of the PES rather than to the stationary one. In this work, our aim is the energy optimization of the phagraphene structure taking into account the possible atomic shifts in the direction perpendicular to the monolayer plane. It is shown that the wavy buckling of the monolayer leads to the lowering of its energy. In addition, we determine the main types of defects arising in phagraphene on heating and determine the characteristic times of their formation.

For the calculations, we use the nonorthogonal tight-binding model [17], which is less accurate than ab initio approaches but demands not so much computer resources and provides an opportunity to study in detail the PES for systems consisting of 300–400 atoms in a quite reasonable time. It also allows analyzing their temporal evolution by the molecular dynamics method. In contrast to the empirical models with the classical interatomic potential, it explicitly takes into account the quantum-mechanical contribution of the electron subsystem to the total energy. The binding energies of different carbon nanostructures and of the bulk forms of carbon determined in the framework of this model are in good agreement with experimental data and ab initio calculations (see [17–20] and references therein). It gives $E_b = -7.35$ eV/atom for graphene and $E_b = -7.14$ eV/atom for phagraphene; i.e., $\Delta E_b = 0.21$ eV/atom, in agreement with [14].

We modeled planar phagraphene using rectangular supercells consisting of $n \times n$ ($n$=1–5) primitive 20-atomic cells (see Fig. 1) with the periodic boundary conditions along the $X$ and $Y$ directions in the plane of monolayer. The structural optimization of each supercell with respect to their periods and the positions of all atoms (at the fixed transverse coordinate $Z = 0$) results in the values of the binding energy $E_b = -7.167, -7.137, -7.138, -7.139$, and $-7.139$ eV/atom at $n = 1, 2, 3, 4$, and 5, respectively. We calculated the normal vibration spectra by the diagonalization of the corresponding Hessians. If the atomic displacements are allowed to occur in the $XY$ plane only, all $2N$ frequencies turn out to be real ($N = 20n^2$ is the number of atoms in the supercell), in agreement with the data reported in [14]. However, account for atomic displacements along the Z axis gives rise to imaginary frequencies in the spectrum the number of which increases from one at $n = 1$ to 33 at $n = 5$. This implies that the planar supercells correspond to saddle points of PES rather than to local minima. Hence, there exist atomic configurations with lower energies.

By analyzing the eigenvectors of Hessians corresponding to imaginary frequencies, we found such configurations at all values of $n$. For each of them, the profile of the transverse (in the XZ plane) atomic displacements has a wavy shape with the wavelength $\lambda$, which can be different at a given $n$ value. Hence, the number $M$ of periods per supercell can be different. At $n = 2$, in particular, there exist two nonplanar low-energy configurations: with $M = 1$ ($\lambda \approx 16$ Å) and $M = 2$ ($\lambda \approx 8$ Å); see Fig. 2. After optimizing the structure with respect to the periods and atomic positions, we found that the minimum energy $E_b = -7.146$ eV/atom corresponds to the configuration with $M = 1$. The energy of the configuration with $M = 2$ is higher by 0.007 eV/atom, whereas it is 0.002 eV/atom lower than that for the planar supercell. For atoms in the supercell, the difference between the maximum and minimum values of Z (the wave amplitude) is $\Delta Z \approx 2.7$ Å at $M = 1$ and $\Delta Z \approx 0.7$ Å at $M = 2$.

For the 3×3 supercell, we found only one nonplanar isomer (with $M = 2$); see Fig. 3. In this case, we have $\lambda \approx 12$ Å and $\Delta Z \approx 1$ Å, and the energy is 0.003 eV/atom lower than that for the planar configuration. The 4×4 supercell has two nonplanar isomers: with $M = 2$ ($\lambda \approx 16$ Å, $\Delta Z \approx 3$ Å) and with $M = 4$ ($\lambda \approx 8$ Å, $\Delta Z \approx 0.6$ Å); see Fig. 4. Their energies are 0.01 and 0.002 eV/atom lower than for the planar isomer, respectively. The frequency spectra for all nonplanar configurations do not exhibit imaginary frequencies.

(a)

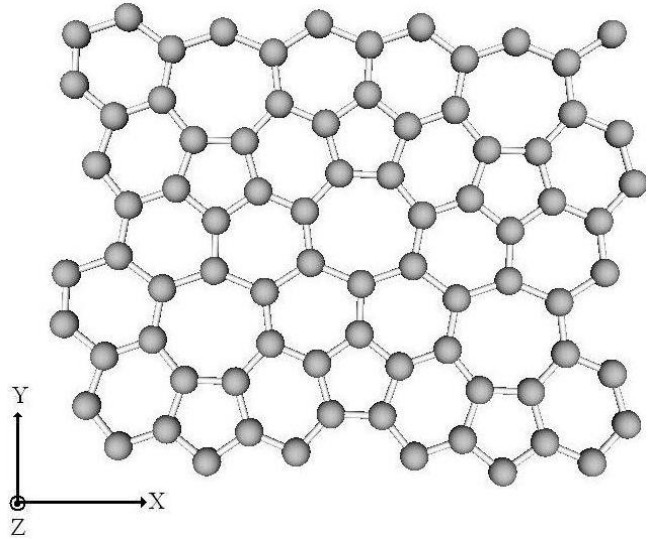

(b)

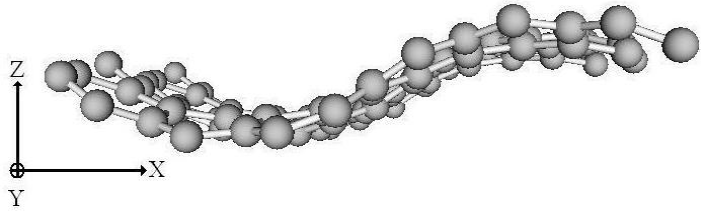

(c)

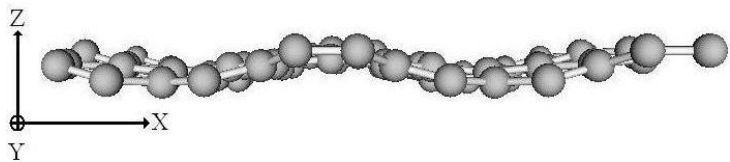

Fig. 2. (a) Planar 2×2 supercell of phagraphene. Side views of its nonplanar configurations with (b) one and (c) two periods of the transverse displacement waves. The binding energies $E_b$ are (a) –7.137, (b) –7.146, and (c) –7.139 eV/atom.

(a)

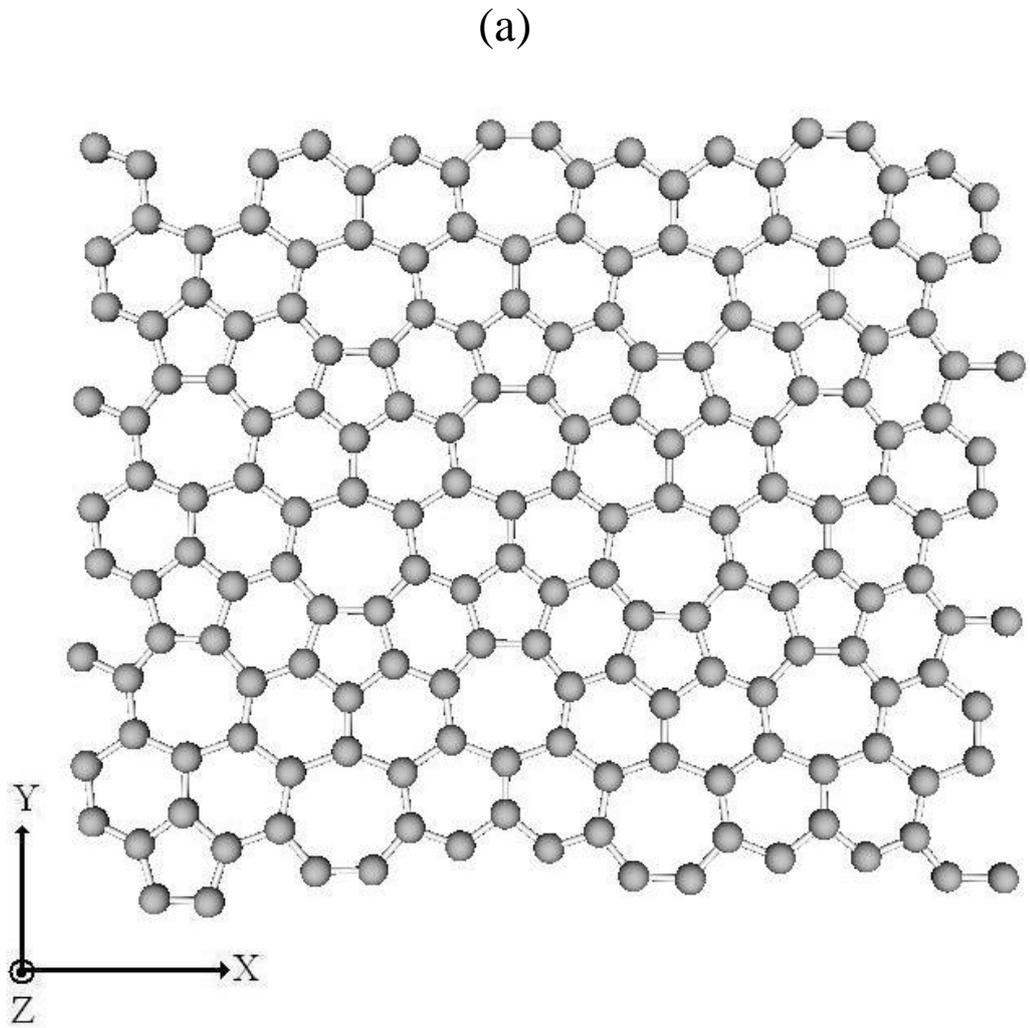

(b)

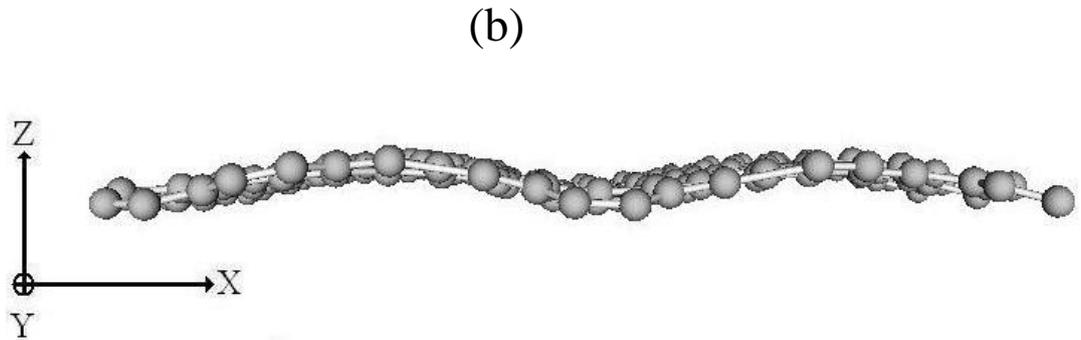

Fig. 3. (a) Planar 3×3 supercell of phagraphene. (b) Side view of its nonplanar configurations with two periods of the transverse displacement waves. The binding energies $E_b$ are (a) –7.138 and (b) –7.141 eV/atom.

(a)

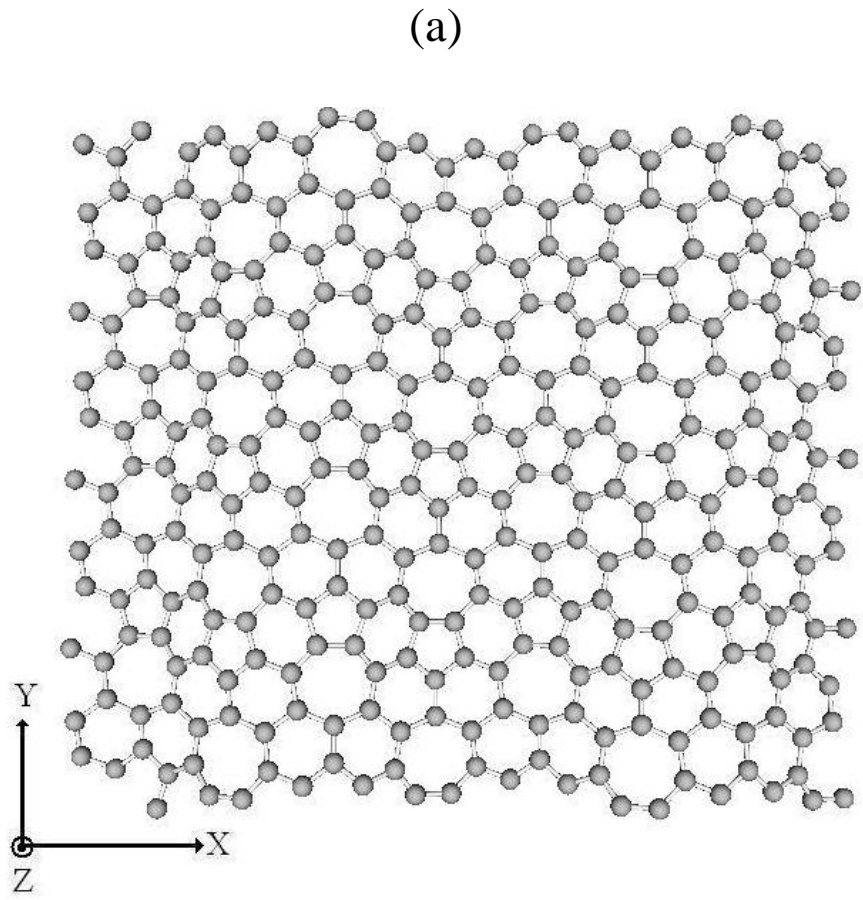

(b)

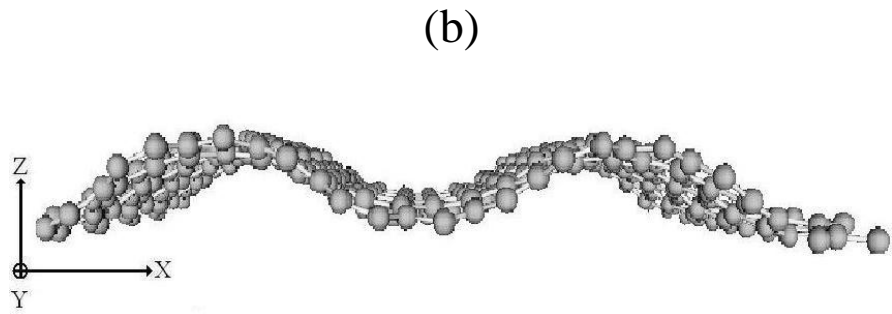

(c)

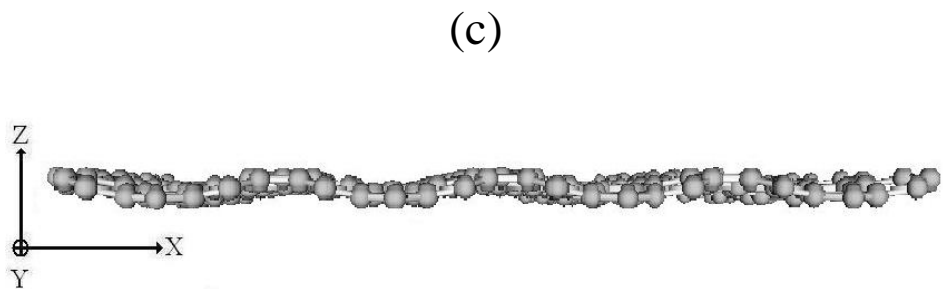

Fig. 4. (a) Planar 4×4 supercell of phagraphene. Side views of its nonplanar configurations with (b) two and (c) four periods of the transverse displacement waves. The binding energies $E_b$ are (a) −7.139, (b) −7.149, and (c) −7.141 eV/atom.

In addition to different phagraphene supercells with periodic boundary conditions, we also studied the structure and energy characteristics of a finite cluster consisting of 50 atoms in which the C–C bonds form neighboring 5-, 6-, and 7-atomic rings (similarly to phagraphene), Fig. 5a. Relaxation of the planar cluster (with free boundary conditions and without the $Z = 0$ limitation) results in its buckling (Fig. 5b), leading to the lowering of its energy by 0.007 eV/atom. The amplitude of the transverse displacements is $\Delta Z \approx 2.6$ Å. Cluster relaxation in the framework of the DFT with the B3LYP/6-311G basis does not change this result qualitatively, although the buckling is not so clearly pronounced ($\Delta Z \approx 0.9$ Å) and the energy drops only by 0.0006 eV/atom.

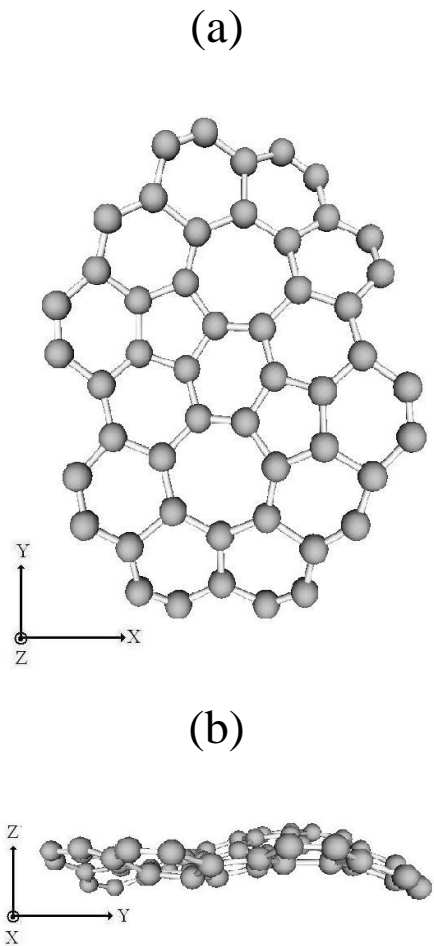

Fig 5. $C_{50}$ cluster with the phagraphene-like structural motif: (a) top view and (b) side view.

Thus, for all phagraphene supercells under study and for the phagraphene-type finite cluster, the non-planar configurations correspond to a lower energy than the planar ones. However, their energies are lower only by $\Delta E \sim 0.001$–$0.01$ eV/atom and the value of $\Delta E$ is quite sensitive to the choice of the calculation method. It is possible that the usage of other calculation methods can lead either to increase in $\Delta E$ or to changes in the order of the alternation of planar and nonplanar configurations on the energy scale. However, regardless of which of these configurations corresponds to the ground state, and which to the excited one, the mere fact of their proximity in energy (near degeneracy) appears to be interesting and important. Indeed, at room (and even at liquid nitrogen) temperature, a small external effect (e.g., deformation) can lead to the transition between these configurations. The coexistence of the planar and nonplanar domains in the same monolayer is also not excluded.

We now discuss the thermal stability of phagraphene. After heating of the 3×3 phagraphene supercell to $T = 1000$ K, the authors of [14] did not find any changes in its structure after 3 ps except for the thermal vibrations of atoms near their equilibrium positions. This is quite natural, because the formation of defects in carbon-based materials with strong covalent bonds requires much longer time and/or higher temperature [20]. We studied the temporal evolution of the 4×4 supercell heated to 4000 K using the molecular dynamics method with a step of about 0.3 fs. After 10 ps, two defects were formed. Each of them was a 8-atomic ring surrounded by five 6-atomic and three 5-atomic rings. Such defects arouse because of the Stone–Wales transformation [15], i.e., the rotation of the C–C bond common to two neighboring 6-atomic rings by an angle of about 90°; see Figs. 6a and 6b. Let us refer to them as defects I. On heating of the 3×3 and 4×4 supercells to different temperatures, $T = (3500–4500)$ K, we also observed the formation of another defects, each being a pair of neighboring 7-atomic rings resulting from the rotation of the C–C bond common to the 6- and 7-atomic rings by an angle of about 90°; see Figs. 6a and 6c. Let us refer to them as defects II. The formation energy (the energy difference between the supercells with and without defects) is $E_d = 0.44$ and $1.39$ eV for defects I and II, respectively. By analyzing the parts of the PES close to the configurations with defects, we found the heights of the energy barriers impeding the formation of defects: $U_I = 4.39$ eV and $U_{II} = 5.06$ eV (recall that the barrier for the formation of the sine-like Stone–Wales defect in graphene is $U \approx 8$ eV [20]). To estimate the characteristic times $\tau_I$ and $\tau_{II}$ for formation of these defects, we use the Arrhenius law $\tau^{-1} = A\exp(-U/k_B T)$, where $A$ is the frequency factor and $k_B$ is the Boltzmann constant. The calculation of the frequency factors according to the Vineyard formula [21] gives $A_I = 1.12 \times 10^{15}$ s$^{-1}$ and $A_{II} = 1.60 \times 10^{14}$ s$^{-1}$. From the Arrhenius law, it follows that times $\tau_I$ and $\tau_{II}$ are macroscopically long at room temperature. However, even after heating to 1500 K, the values of $\tau_I = 0.5$ s and $\tau_{II} = 10$ min are still relatively large. In other words, phagraphene is characterized by a high thermal stability despite the existence of 5- and 7-atomic rings in its structure.

In conclusion, the statement made in [14] concerning the existence of Dirac cones in the electron structure of phagraphene should be revised in view of the results obtained in this work.

We are grateful to M. Maslov for his helpful assistance in performing the calculations and for the discussion of the results. This work was supported by the Russian Foundation for Basic Research, project no. 15-02-02764.

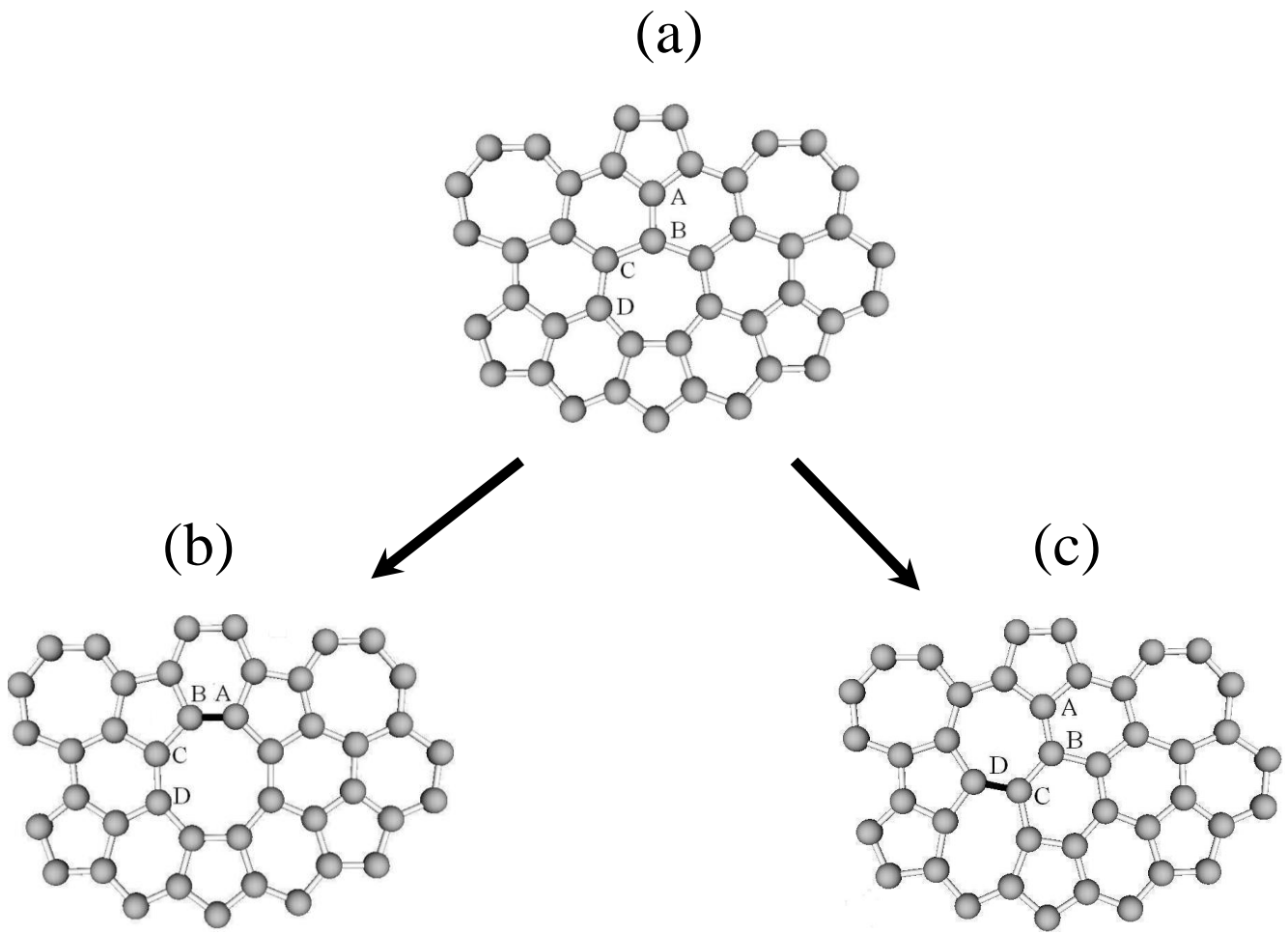

Fig. 6. Illustration of the ways leading to the formation of structural defects in graphene owing to the Stone–Wales transformations: (a) portion of the phagraphene sheet without defects, (b) defect I is formed by the rotation of the AB bond at an angle of about 90°, and (c) defect II is formed by the rotation of the CD bond at an angle of about 90°. The rotated bonds are highlighted in black.

REFERENCES

1. K.S. Novoselov, A.K. Geim, S.V. Morozov, D. Jiang, Y. Zhang, S.V. Dubonos, I.V. Grigorieva, and A.A. Firsov, Science 306, 666 (2004).

2. K.S. Novoselov, A.K. Geim, S.V. Morozov, D. Jiang, M.I. Katsnelson, I.V. Grigorieva, S.V. Dubonos, and A.A. Firsov, Nature 438, 197 (2005).

3. A.E. Galashev and O.R. Rakhmanova, Phys. Usp. 57, 970 (2014).

4. M.M. Haley, Pure Appl. Chem. 80, 519 (2008).

5. L.A. Chernozatonskii, P.B. Sorokin, A.G. Kvashnin, and D.G. Kvashnin, JETP Lett. 90, 134 (2009).

6. G.X. Li, Y.L. Li, H.B. Liu, Y.B. Guo, Y.J. Li, and D.B. Zhu, Chem. Commun. 46, 3256 (2010).

7. A.L. Ivanovskii, Russ. Chem. Rev. 81, 571 (2012).

8. X.-L. Sheng, H.-J. Cui, F. Ye, Q.-B. Yan, Q.-R. Zheng, and G. Su, J. Appl. Phys. 112, 074315 (2012).

9. Y. Liu, G. Wang, Q. Huang, L. Guo, and X. Chen, Phys. Rev. Lett. 108, 225505 (2012).

10. E.A. Belenkov and A.E. Kochengin, Phys. Solid State 57, 2126 (2015).

11. D. Malko, C. Neiss, F. Viñes, and A. Görling, Phys. Rev. Lett. 108, 086804 (2012).

12. H. Huang, W. Duan, and Z. Liu, New J. Phys. 15, 023004 (2013).

13. L.-C. Xu, R.-Z. Wang, M.-S. Miao, X.-L. Wei, Y.-P. Chen, H. Yan, W.-M. Lau, L.-M. Liu, and Y.-M.Ma, Nanoscale 6, 1113 (2014).

14. Z. Wang, X.-F. Zhou, X. Zhang, Q. Zhu, H. Dong, M. Zhao, and A. R. Oganov, Nano Lett. 15, 6182 (2015).

15. A.J. Stone and D.J. Wales, Chem. Phys. Lett. 128, 501 (1986).

16. J. Ma, D. Alfé, A. Michaelides, and E. Wang, Phys. Rev. B 80, 033407 (2009).

17. M.M. Maslov, A.I. Podlivaev, and K.P. Katin, Mol. Simul. 42, 305 (2016).

18. A.I. Podlivaev and L.A. Openov, JETP Lett. 101, 173 (2015).

19. L.A. Openov and A.I. Podlivaev, Physica E 70, 165 (2015).

20. A.I. Podlivaev and L.A. Openov, Phys. Lett. A 379, 1757 (2015) .

21. G.V. Vineyard, J. Phys. Chem. Sol. 3, 121 (1957).